\documentclass[aps,prl,amsmath,amssymb,superscriptaddress,notitlepage,groupedaddress,twocolumn,reprint]{revtex4-1}
\usepackage{chngcntr}\usepackage{graphicx}
\usepackage{dcolumn}
\usepackage{stackengine}
\usepackage[section]{placeins}
\usepackage[dvipsnames*,svgnames]{xcolor}
\usepackage{verbatim}
\usepackage[position=bottom,caption=false,captionskip=3pt,font=normalsize,subrefformat=simple,labelformat=simple]{subfig}

\usepackage{listings}
\usepackage{mathtools}
\usepackage{todonotes}
\usepackage{amsmath}
\usepackage{chngcntr}
\usepackage{bm}
\usepackage{url}

\usepackage[breaklinks]{hyperref}
\usepackage{breakurl}
\usepackage{tabularx}
\usepackage{soul}
\hypersetup{
colorlinks,
linkcolor={blue!100!black},
citecolor={blue},
urlcolor={blue!80!black}}
\newcommand{\fakesection}[1]{\par\refstepcounter{section}\sectionmark{#1}\addcontentsline{toc}{section}{\protect\numberline{\thesection}#1}}\def\equationautorefname~#1\null{#1\null}
\usepackage[bottom]{footmisc} \makeatletter \def\p@section{} \def\p@subsubsection{} \makeatother
\begin{document}
\title{High-Throughput Calculations of Thermal Conductivity in Nanoporous Materials: The Case of Half-Heusler Compounds}
\author{Giuseppe Romano}
\email{romanog@mit.edu}
\affiliation{Department of Mechanical Engineering, Massachusetts Institute of Technology, 77 Massachusetts Avenue, Cambridge, MA 02139, USA}
\author{Jes\'us Carrete}
\affiliation{Institute of Materials Chemistry, TU Wien, A-1060 Vienna, Austria}
\author{David Broido}
\affiliation{Department of Physics, Boston College, Chestnut Hill, Massachusetts 02467, USA}
\author{Alexie M. Kolpak}
\affiliation{Department of Mechanical Engineering, Massachusetts Institute of Technology, 77 Massachusetts Avenue, Cambridge, MA 02139, USA}
\begin{abstract}
Achieving low thermal conductivity and good electrical properties is a crucial condition for thermal energy harvesting materials. Nanostructuring offers a very powerful tool to address both requirements: in nanostructured materials, boundaries preferentially scatter phonons compared to electrons. The search for low-thermal-conductivity nanostructures is typically limited to materials with simple crystal structures, such as silicon, because of the complexity arising from modeling
        branch- and wave vector-dependent nanoscale heat transport. Using the phonon mean-free-path (MFP) dependent Boltzmann transport equation, a model that overcomes this limitation, we compute thermal transport in 75 nanoporous half-Heusler compounds for different pore sizes. We demonstrate that the optimization of thermal transport in nanostructures should take into account both bulk thermal properties and geometry-dependent phonon suppression, two aspects that are typically engineered
        separately. In fact, our work predicts that, given a set of bulk materials and a system geometry, the ordering of the thermal conductivity of the nanostructure does not necessarily align with that of the bulk: We show that what dictates thermal transport is the interplay between the bulk MFP distribution and the nanostructuring length scale of the material. Finally, we derive a thermal transport model that enables fast systems screening within large bulk material repositories and a given geometry. Our study motivates the need for a holistic approach to engineering thermal transport and provides a method for high-throughput materials discovery.\end{abstract}
\maketitle

\fakesection{fake}
 
\section*{Introduction}
 Direct conversion of thermal energy into electricity has tremendous advantages in many applications, including power generation and cooling~\cite{CRC1995}. Despite decades of research on thermoelectric materials, the energy conversion efficiency is still relatively low compared to traditional technologies. The thermoelectric efficiency in semiconductors is limited by the figure of merit $ZT=T \sigma S^2/\kappa$, where $\sigma$ is the electrical conductivity, $S$ is the
        Seebeck coefficient, $T$ the lattice temperature and $\kappa$ is the thermal conductivity (TC), which consists of electronic and lattice components: $\kappa_{el}$ and $\kappa_{L}$. As these three quantities are interrelated, achieving high-$ZT$ materials is challenging. Nanostructuring is a unique platform to overcome some of these challenges because it preferentially suppresses phonon transport relative to electrical transport~\cite{Vineis}. The reason for such behaviour
        stems from the fact that phonon mean free paths (MFPs) are typically larger than electron MFPs that contribute to $\sigma$ in heavily-doped semiconductors. Promising results have been obtained with nanowires~\cite{hochbaum2008enhanced,boukai2008silicon}, thin films~\cite{Venkatasubramanian2001} and porous materials~\cite{marconnet2012phonon,lee2015ballistic,Tang2010,song2004thermal}. As phonons may have wide MFP distributions, effective suppression can be achieved with all-scale hierarchical materials. The different scales can be spanned by combining doping, nano-inclusion and grain engineering~\cite{biswas2012high}.\par 
For practical reasons, the search for low TC bulk materials is often pursued separately from the engineering of phonon suppressing nanostructures. The implicit assumption is that the ordering of bulk TCs within a given material set resembles the relative ordering of TCs in nanostructures based on the same set of materials. In this work, we challenge this assumption by calculating the TC of porous materials based on half-Heusler (HH) compounds. Our model, based on the phonon Boltzmann
        transport equation (BTE) and first-principles calculations, predicts that the bulk ordering is largely preserved only for structures with nanostructuring length scales that are relatively large with respect to the bulk MFP distribution. Conversely, when heat is primarily ballistic, the bulk MFP distribution plays a crucial role in determining the ordering of the TCs. The nanostructuring length scale, referred to as ``characteristic length'' throughout the text, is the limiting dimension
        of the material, e.g. the thickness of a thin film. Finally, we identify a material-independent model, based on a simplified version of the BTE, that provides a faster estimation of the TC of a given nanostructure and a generic set of materials. As HHs are promising thermoelectric materials~\cite{yan2010enhanced,sakurada2005effect}, our work provides practical guidance to experimentalists. Furthermore, it can serve as a base for high-throughput thermal transport in nanostructures, where the bulk MFP distribution can be estimated from either first-principles or experimental reconstructions of MFP distributions~\cite{minnich2011thermal,regner2013broadband}. Finally, it demonstrates that effective material optimization should explicitly include both intrinsic materials properties and system geometries. 
\section*{Results}
\begin{figure}[!ht]

\subfloat[\label{Fig:10a}]
{\includegraphics[width=0.48\columnwidth]{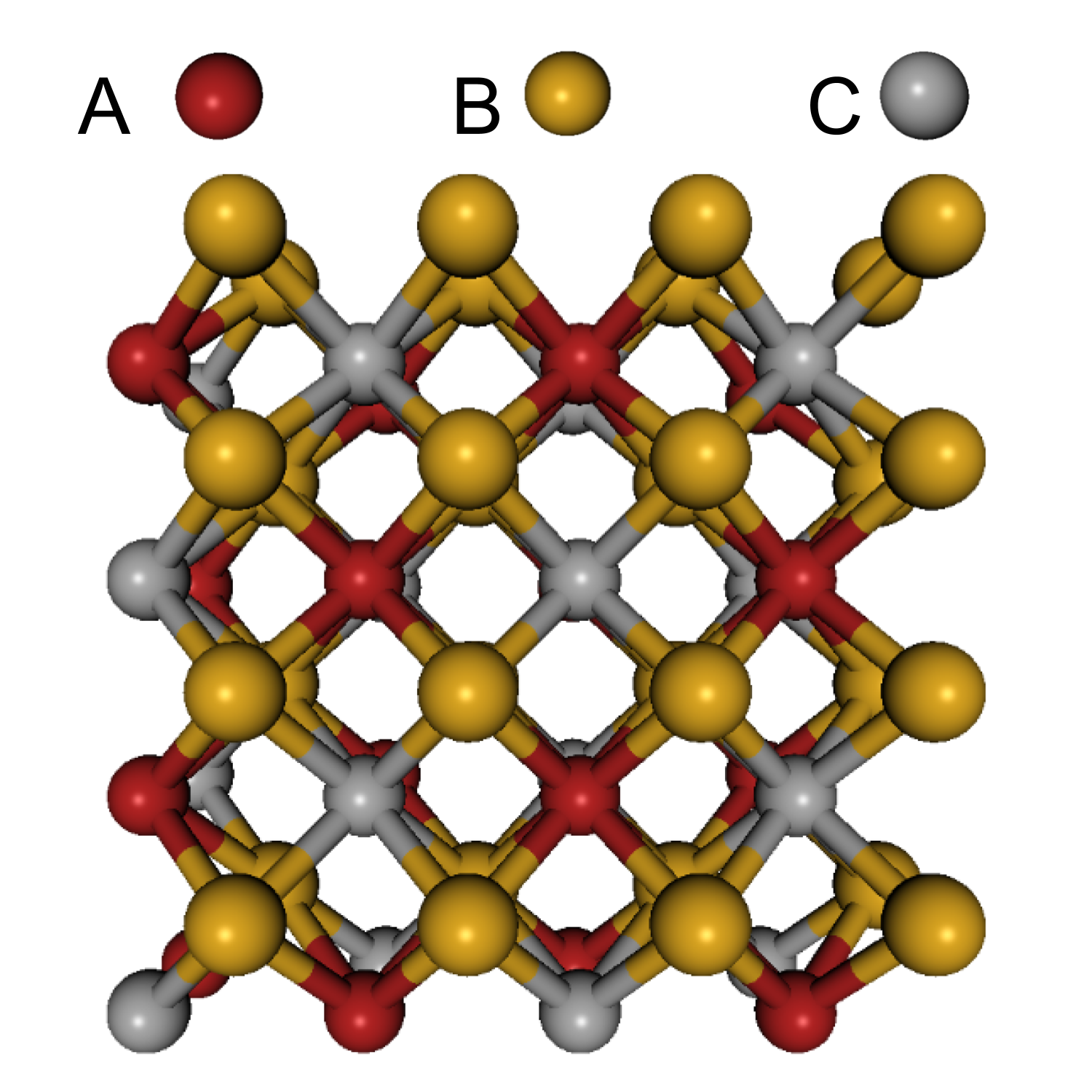}}
\hfill
\subfloat[\label{Fig:10b}]
{\includegraphics[width=0.48\columnwidth]{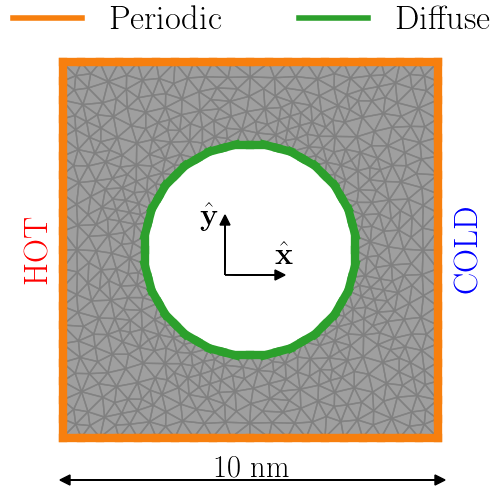}}
\caption{(a) Structure of the HH compound. ABC are three generic elements sitting on three interpenetrating FCC lattices. (b) Unit cell used for the BTE calculation. Heat flux is ensured by imposing a difference of temperature between the hot and cold contacts.}
\end{figure}

\begin{figure*}

\subfloat[\label{Fig:20a}]
{\stackinset{l}{0.25\textwidth}{b}{0.09\textwidth}{\includegraphics[width=0.19\textwidth]{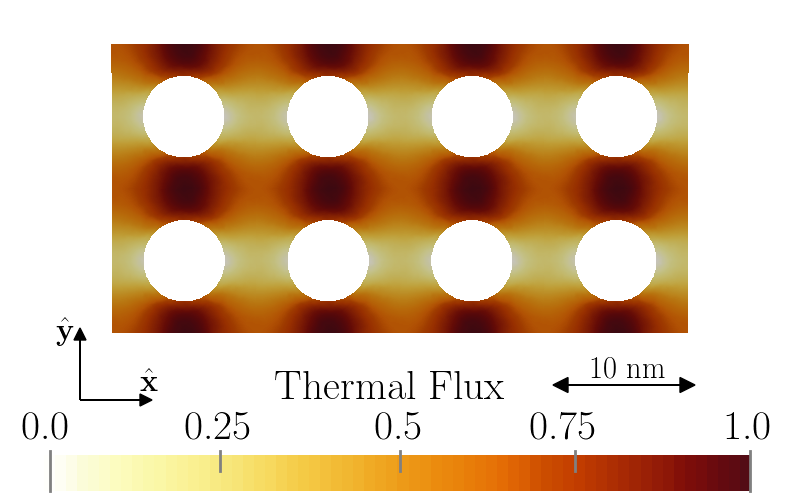}}{\includegraphics[width=0.48\textwidth]{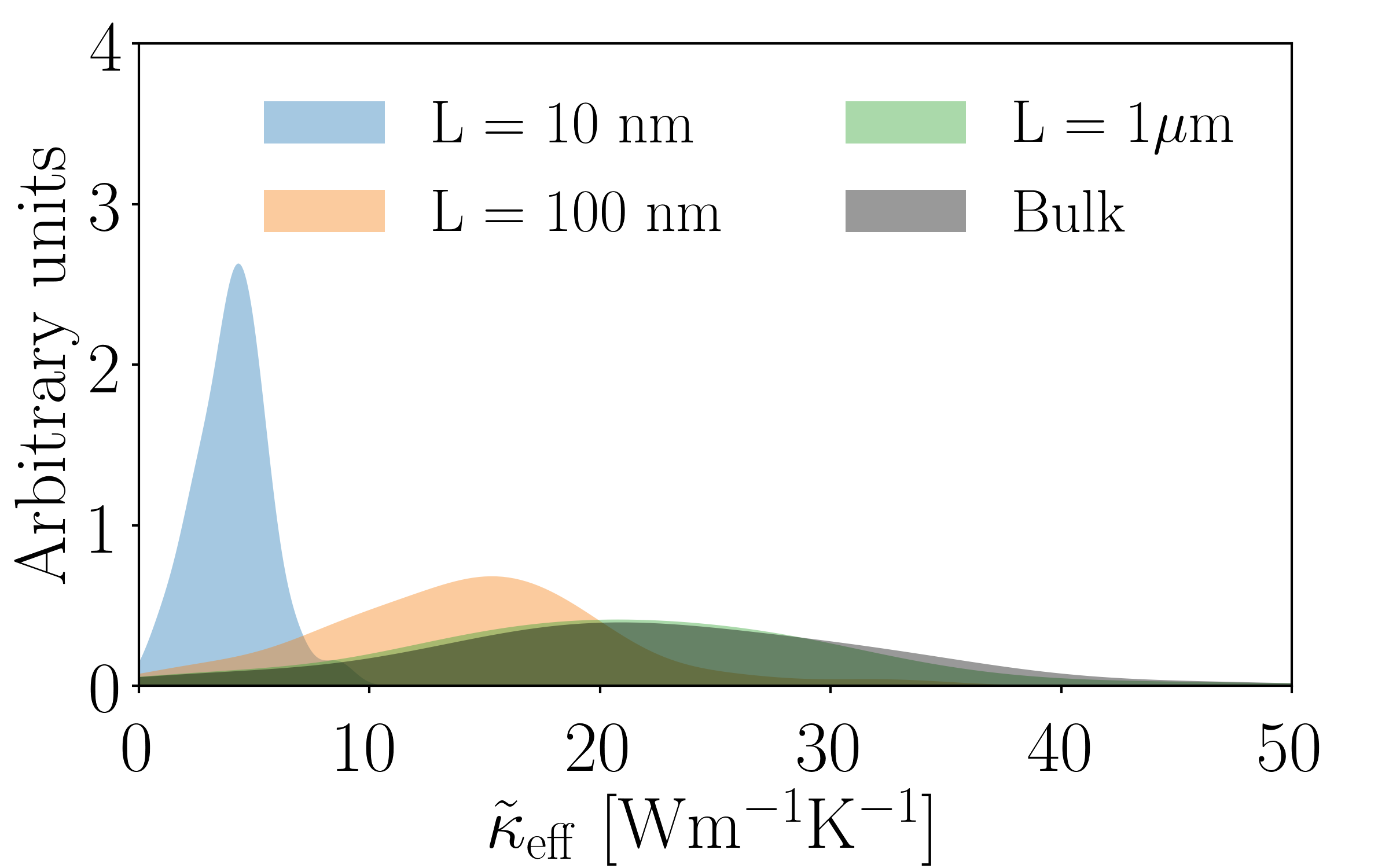}}}
\hfill
\subfloat[\label{Fig:20b}]
{\includegraphics[width=0.48\textwidth]{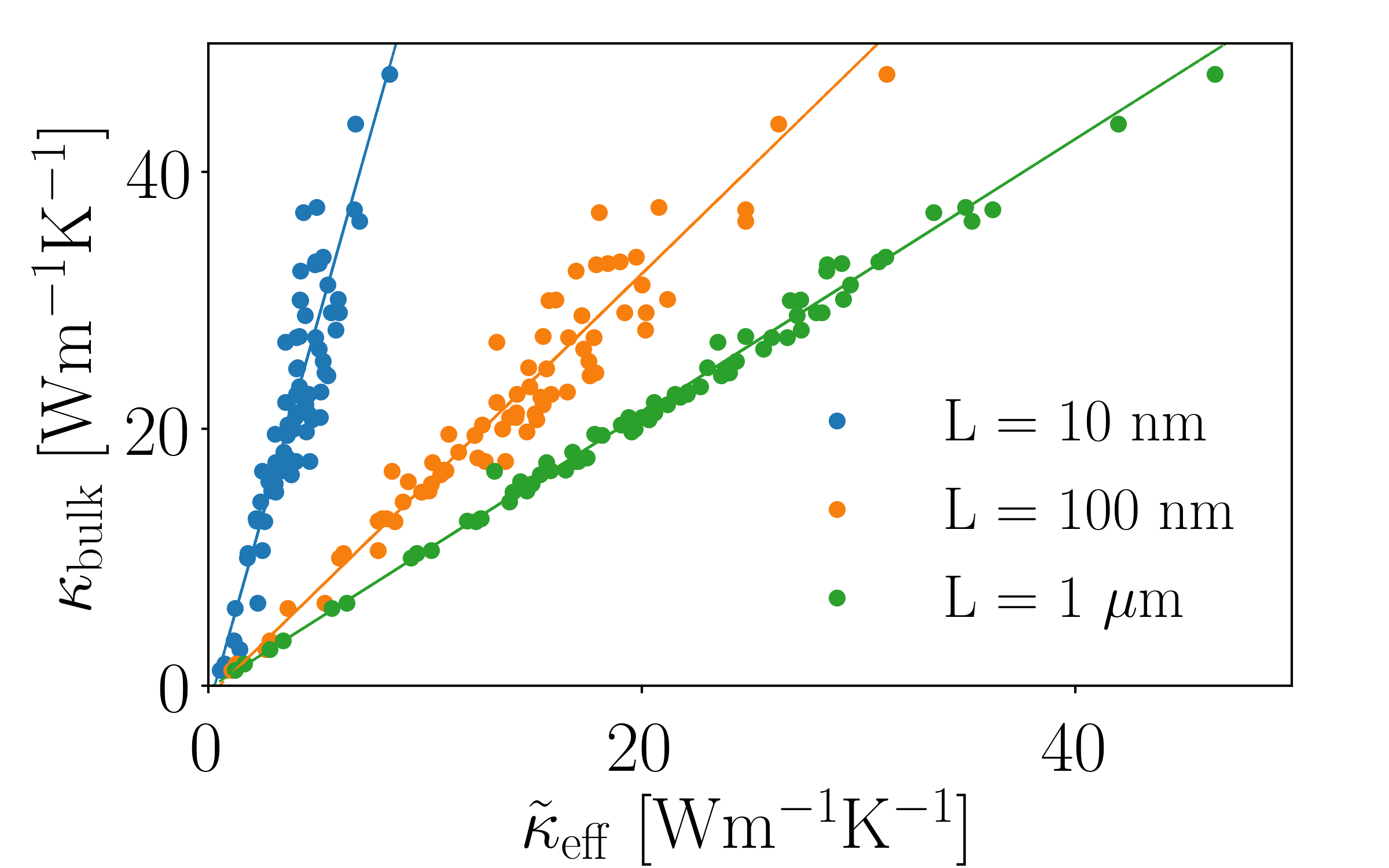}}
\quad
\subfloat[\label{Fig:20c}]
{\stackinset{l}{0.08\textwidth}{b}{0.15\textwidth}{\includegraphics[width=0.18\textwidth]{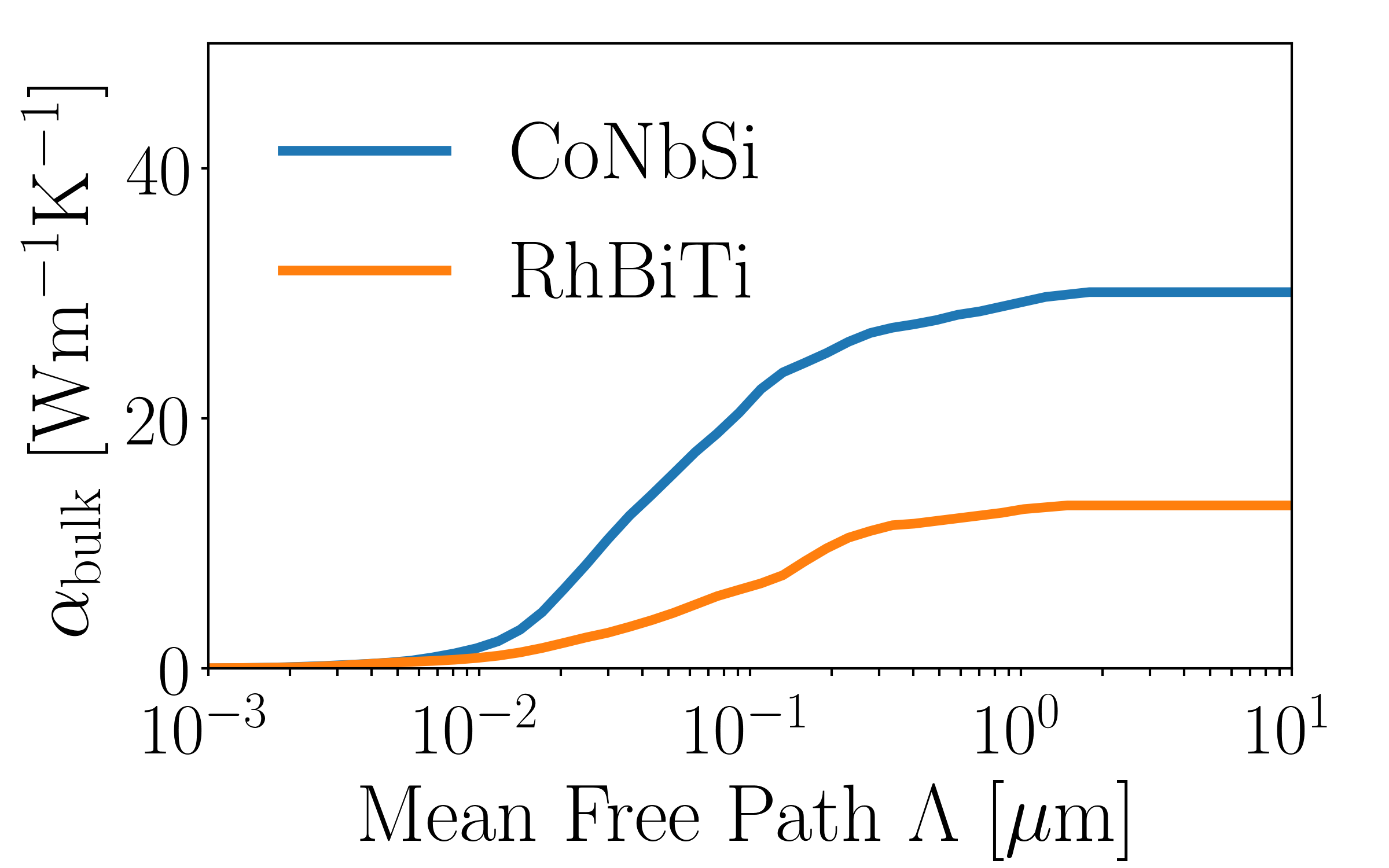}}{\includegraphics[width=0.48\textwidth]{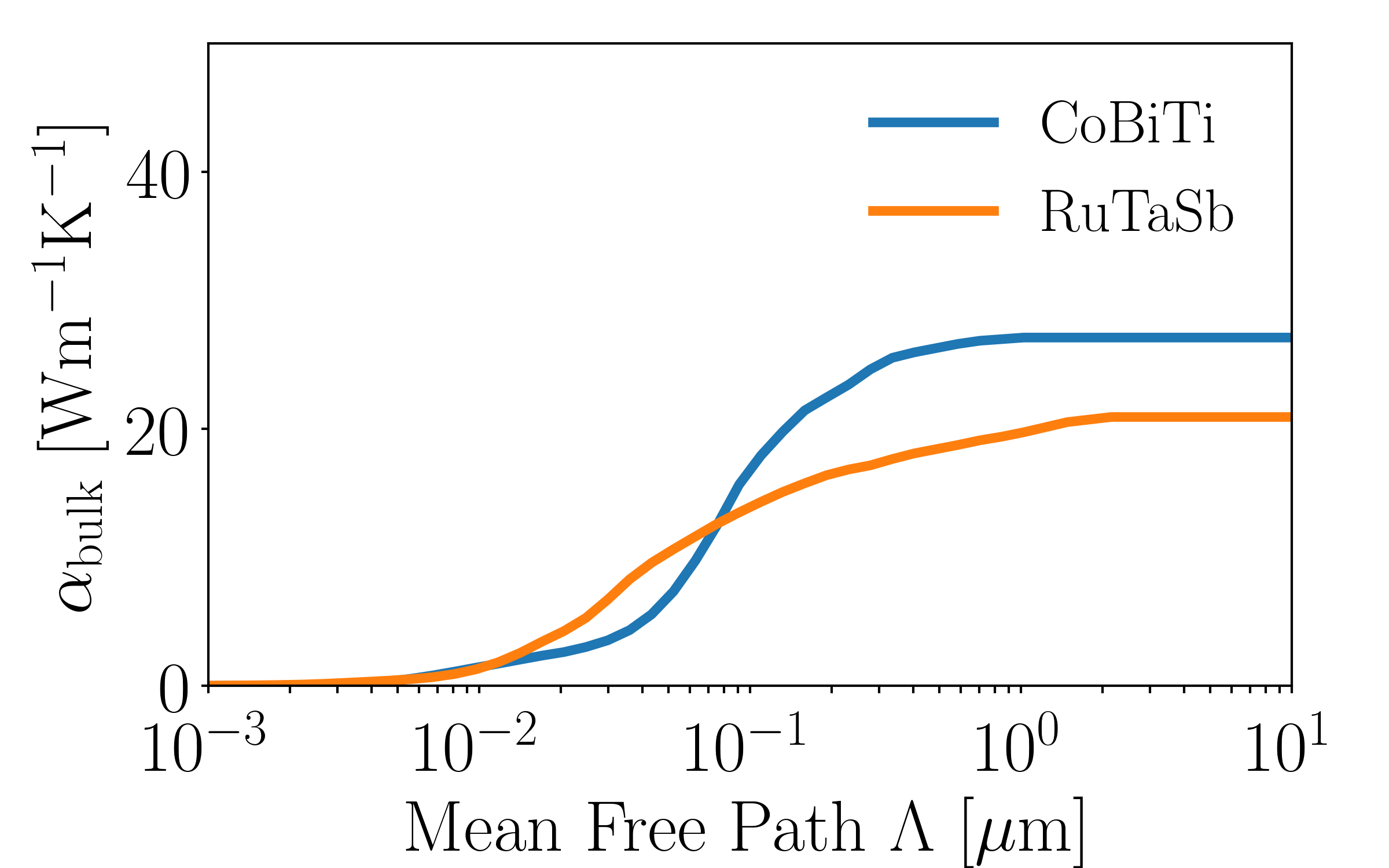}}}
\hfill
\subfloat[\label{Fig:20d}]
{\includegraphics[width=0.48\textwidth]{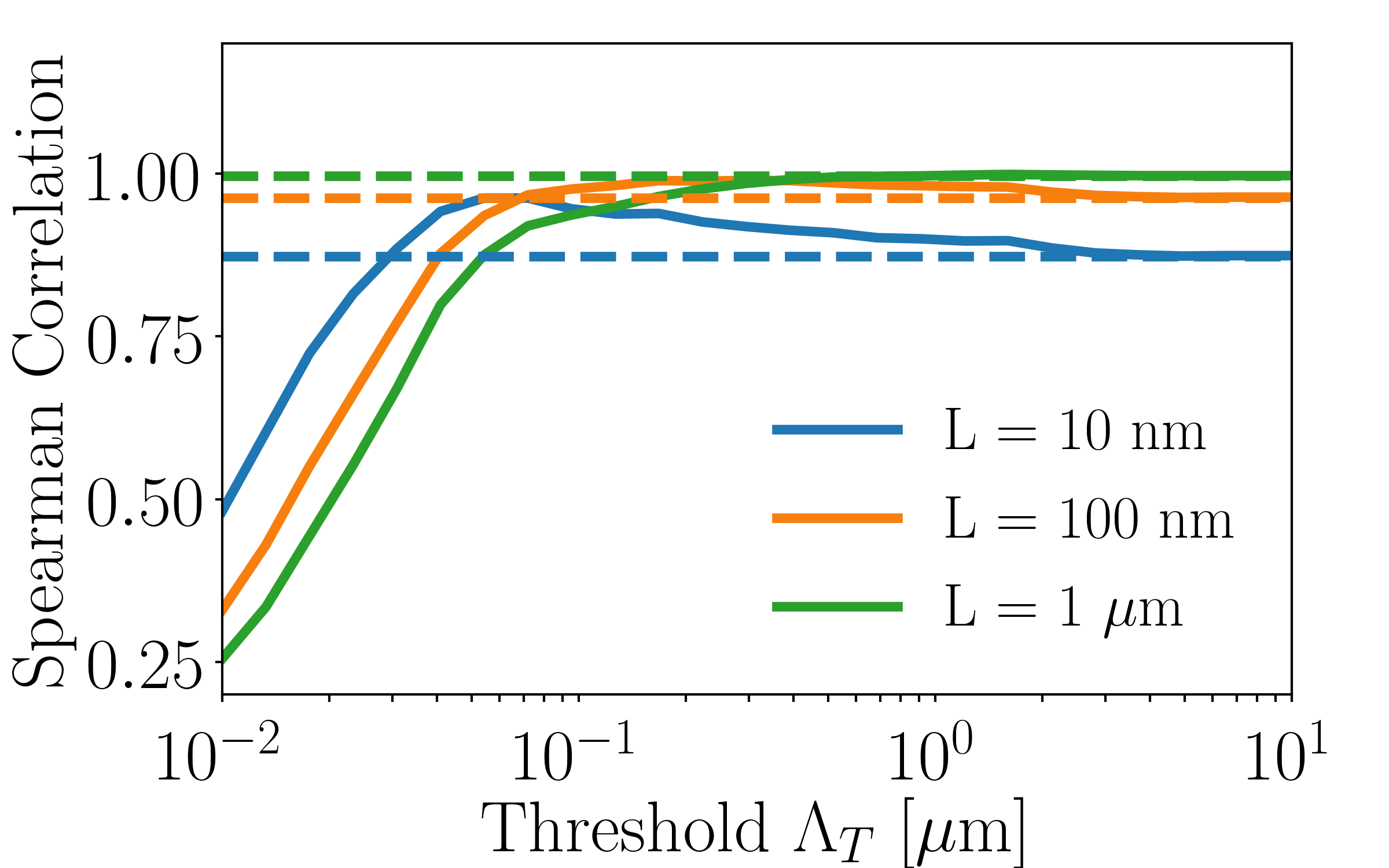}}
\caption{(a) Thermal conductivity ($ \tilde{\kappa}_{\mathrm{eff}} $) density distributions for L = 10 nm, 100 nm and 1 $\mu$m. As the periodicity increases the values of $ \tilde{\kappa}_{\mathrm{eff}} $ approach those of bulk because size effects become negligible. The inset shows the thermal flux map. The gradient of temperature is enforced along the $ \mathbf{\hat{x}} $ direction. Red areas indicate high-flux regions. (b) Linear regression between $ \tilde{\kappa}_{\mathrm{eff}} $ and $ \kappa_{\mathrm{bulk}} $ for all the periodicities. (c) A representive pair of crossing bulk cumulative thermal conductivities, $ \kappa_{\mathrm{bulk}} $. In the inset, a pair of non-crossing $ \alpha_{\mathrm{bulk}} $ is illustrated. (d) The Spearman rank correlation between $ \kappa_\delta(\Lambda_T) $ and $ \tilde{\kappa}_{\mathrm{eff}} $ is shown for all L and varying $\Lambda_T$.}
\end{figure*}

 The crystal structure of the HH compound ABC is formed by three interpenetreting FCC lattices, where A and C form a rocksalt structure and B is located at the diagonal position (1/4, 1/4, 1/4), as shown in Fig.~\subref*{Fig:10a}. The 75 HH compounds considered in this work are taken from Ref.~\cite{PhysRevX.4.011019}, in which high-throughput calculations were used to screen nearly 80 thousand entries from the AFLOW database~\cite{curtarolo2012aflow}, on the basis of mechanical and thermodynamic stability. Similarly to Refs.~\cite{Romano2012he}, the simulation domain of the porous material is a square unit cell comprising a single pore. Periodic boundary conditions are applied to the heat flux on the boundaries of the unit cell. The porosity, \textit{i.e.}, the ratio the pore area to the total area, is fixed to $\phi$ = 0.3, while the size of the unit cell (or the periodicity) is $L$ = 10
        nm, 100 nm, or 1 $\mu$m. Along the walls of the pore we apply diffuse scattering boundary conditions, i.e. \textit{i.e.}, incoming phonons are scattered back isotropically; the temperature of the phonons leaving the pore's surface is set so that zero normal thermal flux is guaranteed along the boundary ~\cite{romano2017hh}. Heat flux is ensured by applying a difference of temperature $\Delta T$ = 1 K between the hot and cold contacts. A sketch of the simulation domain is
        shown in Fig.~\subref*{Fig:10a}. Our model for phonon transport is based on the MFP-BTE under the relaxation time approximation~\cite{romano2011multiscale,romano2015}  \begin{equation}
\begin{split}\label{Eq:1}
 \Lambda \mathbf{\hat{s}} \cdot \nabla T(\mathbf{r},\Lambda) + T(\mathbf{r},\Lambda) = T_L(\mathbf{r}), 
\end{split}
\end{equation}
where $ T(\mathbf{r})$ is the space-dependent effective temperature distribution of phonons with MFP $\Lambda$ and group velocity with direction $ \mathbf{\hat{s}} $; $T_L(\mathbf{r})$ is the effective lattice temperature, given by $T_L(\mathbf{r}) = \int_0^{\infty} <T(\mathbf{r},\Lambda')> a(\Lambda') d\Lambda' $, where $<.>$ is an angular average and $a(\Lambda')= \left[\int_0^{\infty} K_{\mathrm{bulk}} (\Lambda'')/{\Lambda''}^2 d\Lambda'' \right]^{-1} K_{\mathrm{bulk}} (\Lambda')/{\Lambda'}^2 $. The term $ K_{\mathrm{bulk}}
(\Lambda)$ is the bulk MFP distribution, computed by combining density functional theory (DFT) with the phonon supercell approach~\cite{Li2014fg,broido2007intrinsic}. The implementation of our BTE model is described elsewhere~\cite{romano2011multiscale,romano2015}. The essence of classical size effects is captured by the phonon suppression function, $S(\Lambda)$, which describes the ratio of the MFP distribution of the porous material to that of the bulk $ K_{\mathrm{bulk}} (\Lambda)$, where $\Lambda$ is the bulk MFP. Once $S(\Lambda)$ is computed by the BTE, the effective TC is obtained via  \begin{equation}
\begin{split}\label{Eq:2}
\kappa_{eff}=\int_0^\infty K_{\mathrm{bulk}} (\Lambda) S(\Lambda) d\Lambda = \\ =\int_0^\infty \alpha_{\mathrm{bulk}} (\Lambda) g(\Lambda) d\Lambda, 
\end{split}
\end{equation}
where $g(\Lambda) = -\partial S(\Lambda)/\partial \Lambda $ and $ \alpha_{\mathrm{bulk}} (\Lambda) $ is the cumulative bulk thermal conductivity at MFP $\Lambda$, \textit{i.e.} the sum of all MFP contributions up to $\Lambda$. We note the bulk TC is $ \kappa_{\mathrm{bulk}} = \int_0^{\infty} K_{\mathrm{bulk}} (\Lambda)d\Lambda $. The values for $ \kappa_{\mathrm{bulk}} $ at room temperature of the HH compounds considered in this study range from 1.24 WK$^{-1}$m$^{-1}$ (NiHfSn) to 62.12 WK$^{-1}$m$^{-1}$ (CoNbSn)~\cite{PhysRevX.4.011019}. When the porous material is large enough that all phonons travel
        diffusively, $ \kappa_{\mathrm{eff}} $ is obtained by Fourier's law. In this case, the only source of reduction in thermal transport is the decreased volume through which phonons travel due to the presence of the pores. The reduction factor, $\kappa_{\mathrm{Fourier}}/ \kappa_{\mathrm{bulk}} \approx r=(1-\phi)/(1+\phi)$ = 0.54, is predicted by Eucken-Garnett theory~\cite{nan1997effective} and is in agreement with that computed by our Finite-Volume (FV) diffusive solver. As diffusive heat conduction does
        not depend on the phonon MFPs, it gives the same result regardless the size of the unit cell, as long as the porosity is kept constant. Conversely, when size effects occur, $ \kappa_{\mathrm{eff}} $ becomes scale-dependent~\cite{Romano2012}(we use ``scale'' and ``periodicity'' interchangebly.) In order to focus on size effects, we scale $ \kappa_{\mathrm{eff}} $ by the macroscopic reduction factor, i.e. $ \tilde{\kappa}_{\mathrm{eff}} = \kappa_{\mathrm{eff}} \kappa_{\mathrm{bulk}} /\kappa_{\mathrm{Fourier}} = \kappa_{\mathrm{eff}} r^{-1}$. In Fig.~\subref*{Fig:20a}, we plot
        the distribution of $ \tilde{\kappa}_{\mathrm{eff}} $ for all compounds and periodicities. We note that for L = 10 nm, most of the values are below 10 Wm$^{-1}$K$^{-1}$, as a result of suppression of long-MFP phonons. As L increases, the distribution of $ \tilde{\kappa}_{\mathrm{eff}} $ widens up until approaching the bulk one for L = 1 $\mu$m, where size effects become negligible. The values of $ \tilde{\kappa}_{\mathrm{eff}} $ are available upon request.\par 
 
We now assess whether $ \kappa_{\mathrm{bulk}} $ can be used as a ``descriptor'' for $ \kappa_{\mathrm{eff}} $. A descriptor is a simple model correalted to, within some approximation, a more complicated calculation. We perform a linear regression between these two quantities, as shown in Fig.~\subref*{Fig:20b}. We quantify their correlation with the Spearman rank correlation, a statistical quantity that indicates the monotonic correlation between two variables~\cite{carrete2014finding}, obtaining $\approx 0.88, 0.96, 1$ for L = 10 nm, 100 nm and 1 $\mu$m, respectively. This trend can be understood if we analyze phonon suppression in terms of $S(\Lambda)$ and $ K_{\mathrm{bulk}} (\Lambda) $, as encoded in Eq.~(\autoref{Eq:2}). As a first approximation, we can assume that heat carried by phonons with $\Lambda$ below a given threshold $\Lambda_T$ does not suffer size effects, while all the rest is completely suppressed. Within this assumption, the suppression function is given by $S(\Lambda) =r \Theta(\Lambda_T - \Lambda)$, where $\Theta(x)$ is the Heaviside function. Using Eq.~(\autoref{Eq:2}), we get $g(\Lambda)=r\delta(\Lambda - \Lambda_T)$ and $ \tilde{\kappa}_{\mathrm{eff}} \approx \tilde{\kappa}_\delta= \alpha_{\mathrm{bulk}}(\Lambda_T)$. This result shows that $ \tilde{\kappa}_{\mathrm{eff}} $ is dictated by the bulk cumulative thermal conductivity around $\Lambda_T$ rather than by $ \kappa_{\mathrm{bulk}} $. In fact, there are cases, such as for the pair CoBiTi and RuTaSb, where the curves of $ \alpha_{\mathrm{bulk}} $ cross each other for some values of $\Lambda_T$ (see Fig.~\subref*{Fig:20c}). In this instance, if the nanostructuring length is smaller than the crossing point the ordering of $ \kappa_{\mathrm{bulk}} $ is the opposite to that of $ \tilde{\kappa}_{\mathrm{eff}} $, an effect that is captured by $ \tilde{\kappa}_\delta $. For completeness, we note that there are cases, such as the pair CoNbSi-RhBiTi (see inset of Fig.~\subref*{Fig:20c}), where the ordering swaping is absent. It is worth noting that the effect of ordering mismatch has already been discussed conceptually in~\cite{romano2016temperature}.\par 
 
\begin{figure*}

\subfloat[\label{Fig:30a}]
{\includegraphics[width=0.48\textwidth]{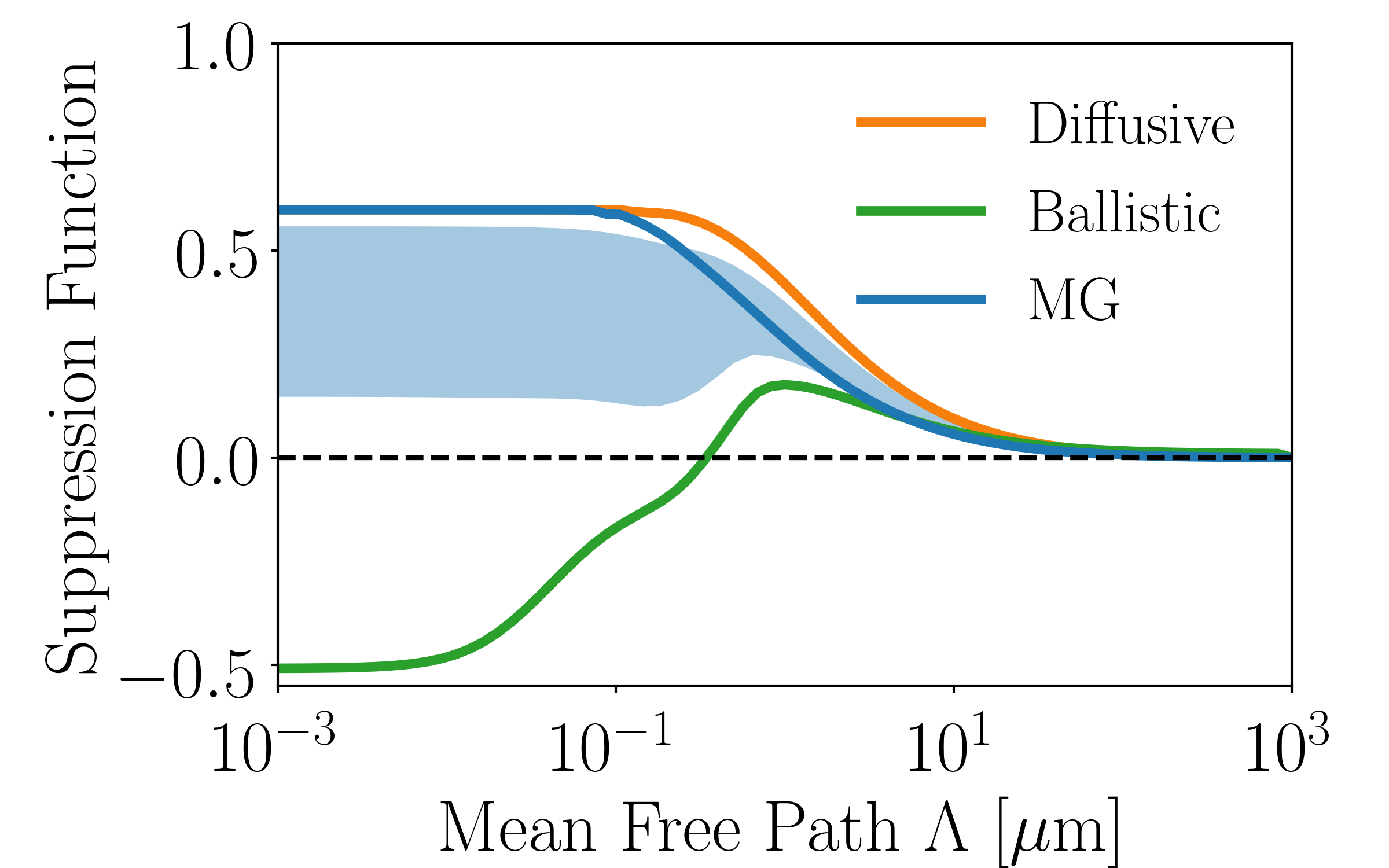}}
\hfill
\subfloat[\label{Fig:30b}]
{\includegraphics[width=0.48\textwidth]{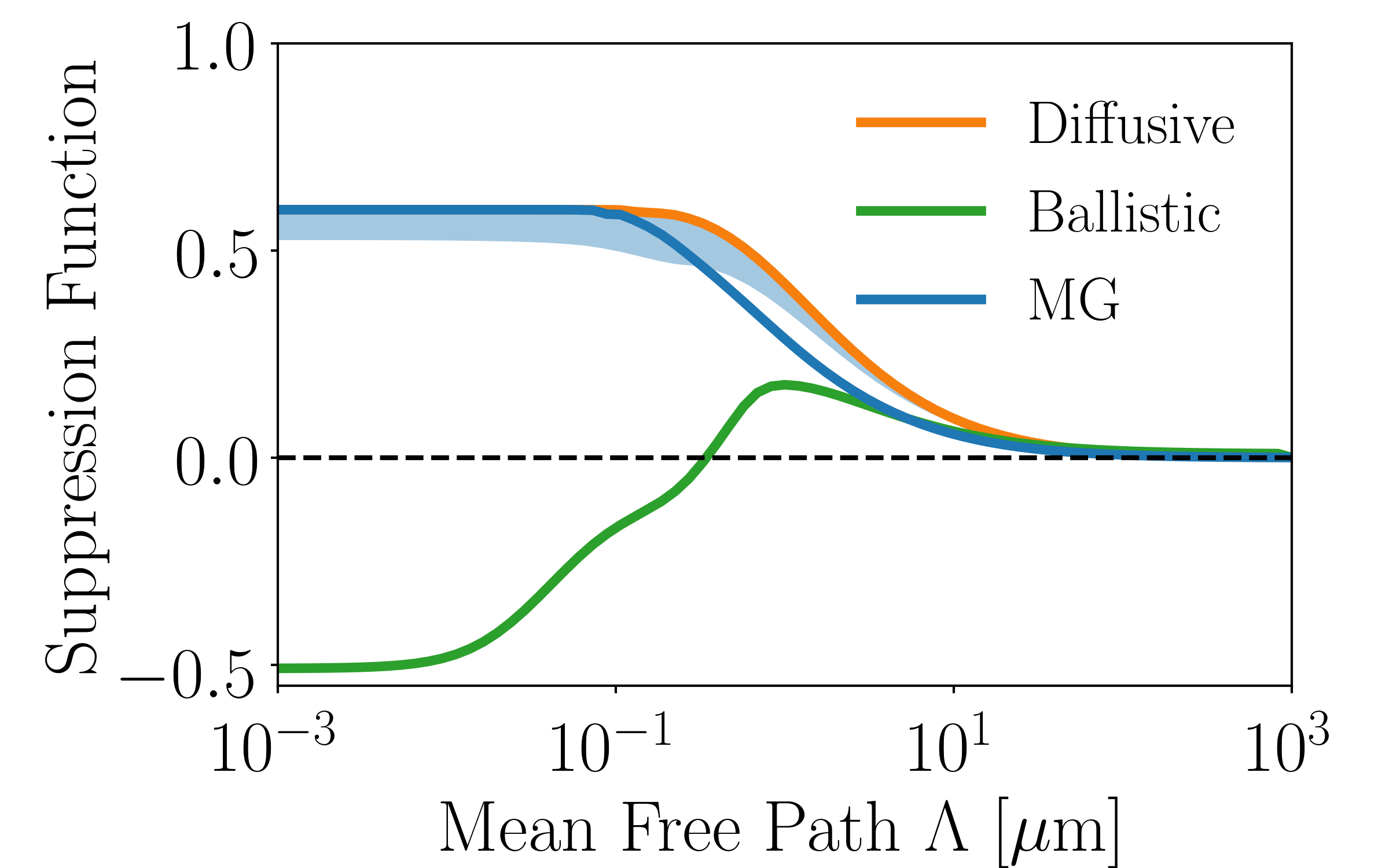}}
\quad
\subfloat[\label{Fig:30c}]
{\stackinset{l}{0.08\textwidth}{b}{0.16\textwidth}{\includegraphics[width=0.17\textwidth]{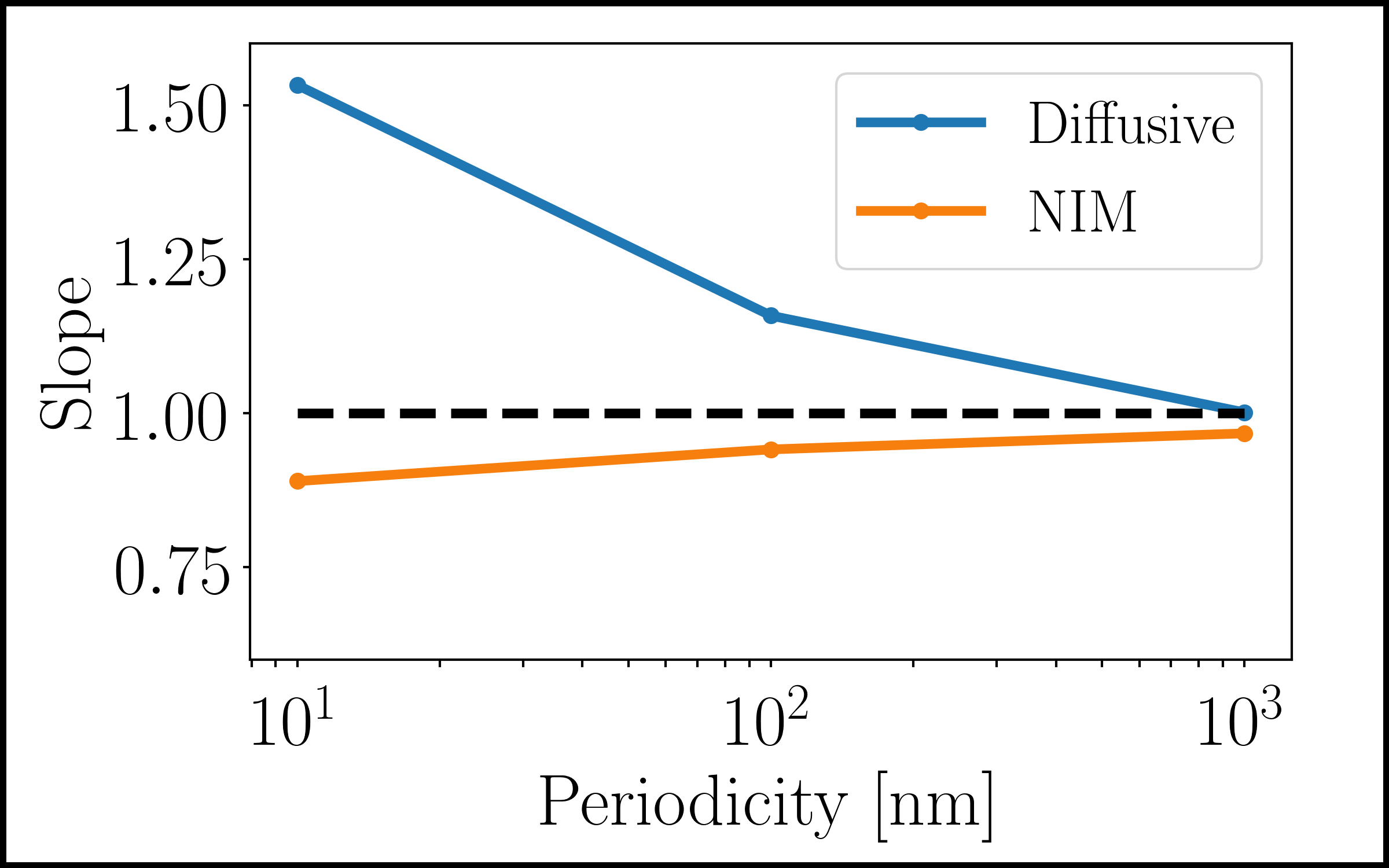}}{\includegraphics[width=0.48\textwidth]{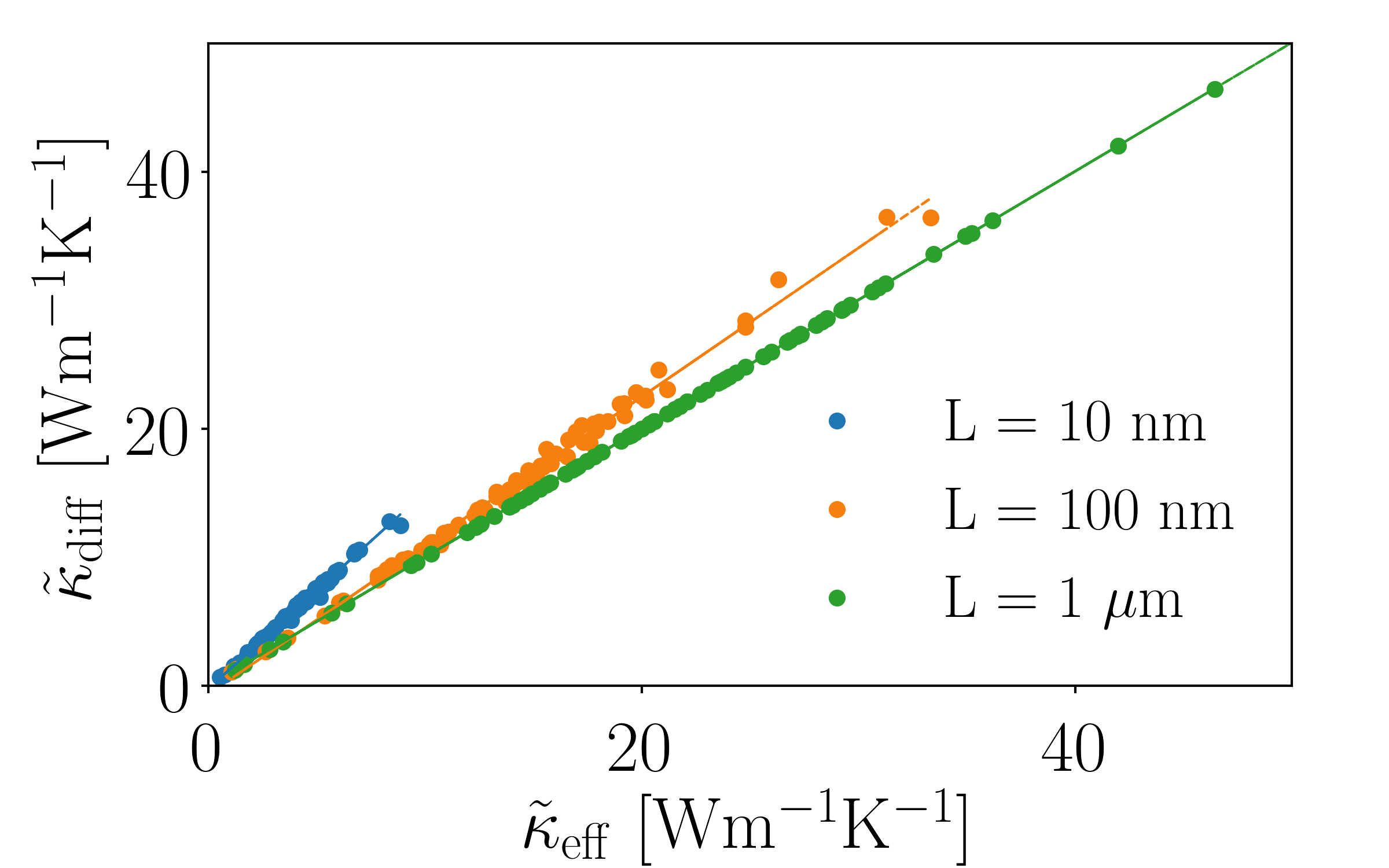}}}
\hfill
\subfloat[\label{Fig:30d}]
{\includegraphics[width=0.48\textwidth]{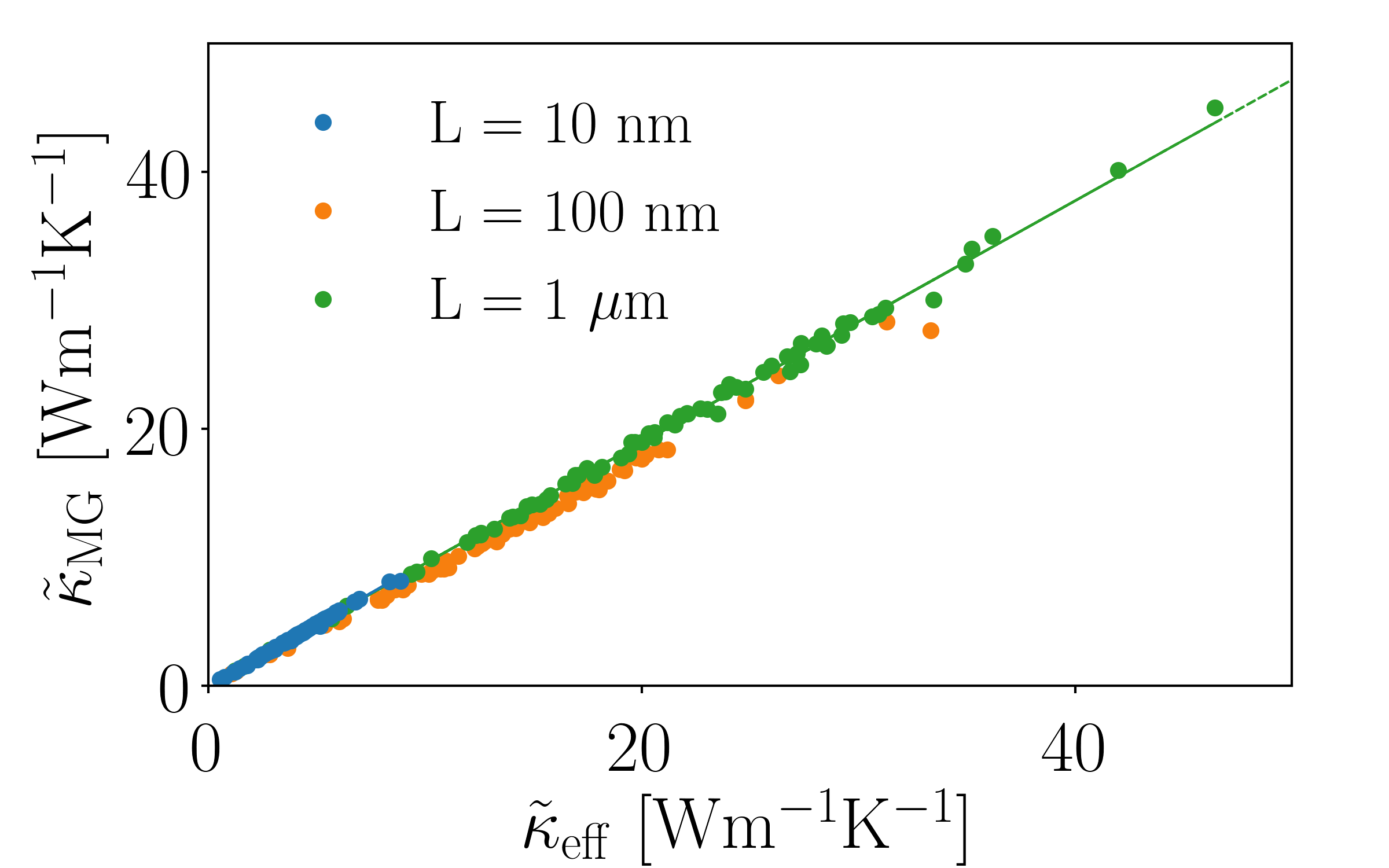}}
\caption{Suppression functions of the HH compounds (blue shaded region) for (a) L = 10 nm and (b) L = 100 nm. These curves are bounded by those of the ``diffusive'' and ``ballistic'' material limits. The suppression function of the ``multiple gray'' model is also plotted. (c) The linear regression between $ \tilde{\kappa}_{\mathrm{diff}} $ and $ \tilde{\kappa}_{\mathrm{eff}} $. In the inset, the slope of the regression models for both $ \tilde{\kappa}_{\mathrm{diff}} $ and $\tilde{\kappa}_{MG}$. (d) The regression model between $ \tilde{\kappa}_{\mathrm{eff}} $ and $ \tilde{\kappa}_{MG}$. }
\end{figure*}
In this section we assess the use of $ \kappa_\delta $ as a descriptor for $ \kappa_{\mathrm{eff}} $.  To this end we calculate the Spearman rank correlation between these two quantities for different $ \Lambda_T $. As shown in Fig.~\subref*{Fig:20d}, for all the periodicities such a correlation increases with $\Lambda_T$, reaches a maximum, and eventually approaches a constant value. These final values concide with those obtained with $ \kappa_{\mathrm{bulk}} $ as a decriptor since
        $\alpha_{\mathrm{bulk}}(\infty) = \kappa_{\mathrm{bulk}} $. The value at which the correlation is maximum increases with $L$, as a consequence of the increasing characteristic length. These results show, therefore, that by taking into account $ \alpha_{\mathrm{bulk}} $ the ordering of $ \tilde{\kappa}_{\mathrm{eff}}  $ within a given set of materials is better estimated with respect to the simple use of $ \kappa_{\mathrm{bulk}} $. However, this new descriptor has a limitation: the optimal $\Lambda_T$ is unknown unless one runs the BTE, negating the utility
        of $ \kappa_\delta $. Motivated by this shortcoming, we introduce a parameter-free descriptor, as described below.\par 
In a previous work~\cite{romano2017hh}, we showed that the suppression function of a given material and geometry is bounded by the ``diffusive'' and ``ballistic'' material approximations. A diffusive (ballistic) material is the case where all the bulk phonon MFPs are larger (smaller) than the characteristic length. The suppression functions of these two limits along with those of the HH compounds are shown in Fig.~\subref*{Fig:30a} and Fig.~\subref*{Fig:30b} for L = 10 nm and 100 nm, respectively. We note that, while in the former case the curves of $S(\Lambda)$ cover a wide range (due to strong size effects), for larger periodicities the suppression functions are compressed toward the diffusive material limit because of weaker size effects. In light of this result, we then speculate whether the suppression function of the diffusive material limit, referred to as $S_{\mathrm{diff}}(\Lambda)$, can be used to estimate $ \kappa_{\mathrm{eff}} $, i.e. $ \tilde{\kappa}_{\mathrm{eff}} \approx \tilde{\kappa}_{\mathrm{diff}} =r^{-1} \int_0^\infty S_{\mathrm{diff}}(\Lambda) K_{\mathrm{bulk}} (\Lambda) d\Lambda $. Interestingly, The Spearman rank correlation is close to unity for all the considered periodicities, corroborating the use of $ \tilde{\kappa}_{\mathrm{dff}} $ as an effective descriptor for $ \tilde{\kappa}_{\mathrm{eff}} $. As $S_{\mathrm{diff}}(\Lambda)$ is scale- and material- independent, it would need to be computed only once, for a particular geometry, and then be used for a generic material and periodicity, increasing dramatically the computational efficiency of material screening.\par 
Although $ \tilde{\kappa}_{\mathrm{diff}} $ estimates the ordering of $ \kappa_{\mathrm{eff}} $, it does not provide a prediction of the absolute values of the effective TC when size effects are significant (see Fig.~\subref*{Fig:30c}). In fact, the slopes of the linear regressions are $~$1.53, 1.16 and 1 for L = 10 nm, 100 nm and 1 $\mu$m, respectively. To overcome this limitation we introduce a model based on solving the BTE for each phonon MFP independently. In practice, this is the equivalent to the ``gray'' model solved for diffent values of $\Lambda$. We will refer to this method as the ``multiple gray'' model (MG). The resulting suppression function, $S_{\mathrm{mg}}(\Lambda)$, is then used to compute $ \tilde{\kappa}_{\mathrm{eff}} \approx \tilde{\kappa}_{\mathrm{MG}} = r^{-1} \int_0^\infty S_{\mathrm{MG}}(\Lambda) K_{\mathrm{bulk}} (\Lambda) d\Lambda $. Within this model, Eq.~(\autoref{Eq:1}) becomes  \begin{equation}
\begin{split}\label{Eq:3}
 \Lambda \mathbf{\hat{s}} \cdot \nabla T(\mathbf{r},\Lambda) + T(\mathbf{r},\Lambda) = < T(\mathbf{r},\Lambda)>. 
\end{split}
\end{equation}
The curve of $ S_{\mathrm{MG}}(\Lambda)$, as shown in Fig.~\subref*{Fig:30a}, is close to the diffusive one; yet, the predicting power of $\tilde{\kappa}_{MG}$ is higher than that of $\tilde{\kappa}_{\mathrm{diff}}$ for most of the length scales, yielding a slope in the regression model, shown in Fig.~\subref*{Fig:30d}, of 0.89, 0.94 and 0.97, for L = 10 nm, 100 nm and 1 $\mu$m, respectively (see inset in Fig.~\subref*{Fig:30c}). In light of these results we recommend using the MG model. \par 

\section*{Conclusion}
 By solving the phonon Boltzmann transport equation, we have computed the effective thermal conductivity of 75 nanoporous half-Heusler compounds with different periodicities, obtaining significant reduction with respect to the bulk. Then, we have developed a model that enables the calculation of thermal transport in a large number of materials by solving the BTE only once, within a given geometry. In addition to enhancing our understanding of nanoscale heat transport, our approach has the potential of accellerating materials discovery for thermoelectric applications. 
\section*{Acknowledgements}
 Research supported as part of the Solid-State Solar-Thermal Energy Conversion Center (S3TEC), an Energy Frontier Research Center funded by the US Department of Energy (DOE), Office of Science, Basic Energy Sciences (BES), under Award DESC0001. 
\end{document}